\documentclass{article}

\begin{document}

\begin{center}
{\LARGE Supermassive galactic centre with repulsive gravity}

Trevor W. Marshall

and

Max K. Wallis

Buckingham Centre for Astrobiology,

The University of Buckingham,

Buckingham MK18 1EG, UK
\end{center}

\bigskip \textbf{Abstract}

Repulsive gravity has its origin in the 1939 article of Oppenheimer and
Snyder which describes a collapsar, that is an idealized star of
noninteracting material (dust) collapsing under its own gravity. The stellar
material has a final state resembling a football, that is a significant part
of it is concentrated in a thin surface shell. An interior pressure is
exerted by the strong gravitational field, equivalent to a negative mass.
However, the OS solution has been misunderstood, the shell's position being
incorrectly identified with the "event horizon" in black-hole theory. While half the material is concentrated in a shell occupying a
small fraction of the radius, some material is spread throughout the
interior, unlike the concentration in a black hole's singularity. We deal
with the singularity in density at the shell surface, by including Fermi
pressure of degenerate electrons for a shell density comparable to a solar
mass-sized white dwarf. Because the high-density region is concentrated in a shell, instead of at the centre as in a black hole, our conclusion is that repulsive gravity enables the existence of supermassive white dwarfs.

\section{\protect\bigskip Introduction}

Repulsive gravity enables us to model supermassive galactic centres such as
Sagittarius A* with most of the material concentrated in a shell which is
inflated by a strong interior gravitational field. The mass density in the
shell is similar to that in white dwarf stars, so we anticipate that
degenerate electron Fermi pressure balances the repulsive gravitational
field and spreads out the singularity in surface density of the dust star
model. We call this a supermassive white-dwarf (SWD). Detailed investigation
of SgrA* at the gravitational radius of OS, as is envisaged in the Event
Horizon Telescope project, should discriminate very clearly between the SWD
and black-hole models, as electromagnetic rays behave very differently in
the two. Accretion of matter by SgrA* from the G2 approaching dust cloud
would also be different in the two models.

The earliest indication that gravity may be repulsive is in the 1939 article
of Oppenheimer and Snyder\cite{oppsny} (OS), describing an idealized
collapsar composed of "dust", that is matter which distorts the
gravitational metric but has no interaction with itself. Penrose\cite{penrose} and
subsequent experts in \ General Relativity (GR) failed to appreciate the
novelty of the OS solution; it is in fact an immediate consequence of the OS
metric that their collapsar has a stable end state whose radius is what they
termed the \emph{gravitational radius -- }nowadays frequently referred to as
the Schwarzschild radius. It is only fair to point out that at least the
better known of the OS authors should be given some part of the blame for
the misunderstanding, since J. R. Oppenheimer had coauthored, with G. M.
Volkoff a few months earlier an article\cite{oppvol} (OV) which tried to
analyse the collapse by applying the field equations of General Relativity (GR) to extend the Newtonian theory of white dwarfs. The conclusion of OV was that any object whose mass exceeds a small multiple of the solar mass must collapse to zero radius. Though the authors of OS failed to acknowledge it, their result contradicted OV. 

OS noted that the speed of light approaches zero at the gravitational radius, where the mass concentrates, so contraction to the end state takes infinite time.  However, one of us previously found\cite{causaps} that a simple modification to the OS metric can be introduced to remove the light speed problem and thereby meet the Hilbert causality condition (ie. connectivity via light rays). The  singularity at the surface is removed by adding a 'gas' pressure term, as used by OV, which spreads out the peak in mass. We are not concerned over the actual form of gas pressure, but we do find that  temperature independent polytropic forms like those used for white dwarf stars (Fermi pressure of degenerate electrons) are simple and amenable to analysis. Such polytropic models were investigated in the 1930s, but OV  used them to exclude wrongly those solutions with zero central pressure which follow from the OS analysis.

\section{The OS metric}

\bigskip \emph{A rock, dug up 74 years ago, is wrongly classified as quartz
and left on a museum shelf. After washing, it is reclassified as a huge
diamond, about a hundred times the Koh-i-Noor.}

The surface of the OS collapsar describes a free-fall geodesic of the
Schwarzschild metric, that is its radius $r_{0}\left( t\right) $ satisfies%
\begin{equation}
t=-\frac{2}{3}r_{0}^{3/2}-2\sqrt{r_{0}}+\ln \frac{\sqrt{r_{0}}+1}{\sqrt{r_{0}%
}-1}\quad ,
\end{equation}%
where we have chosen units in which the gravitational radius $2MG/c^2$ is 1. The
fact that $t$ tends to plus infinity as $r_{0}$ tends to 1 led OS to infer
at the end of their Abstract that "....an external observer sees the star
asymptotically shrinking to its gravitational radius". The interior $r\left(
t\right) <r_{0}\left( t\right) $ is described by the metric%
\begin{equation}
ds^{2}=\frac{r^{3}}{R^{3}}\left( \frac{dr}{r}-\frac{dR}{R}\right) ^{2}-\frac{%
r^{2}}{R^{2}}dR^{2}-r^{2}\left( d\theta ^{2}+\sin ^{2}\theta d\phi
^{2}\right) \quad ,
\end{equation}%
where $R$, with $0<R<1,$ is a \emph{comoving coordinate, }meaning that the
interior free-fall geodesics are given by $R=$constant; in particular $R=1$
is identical with $r=r_{0}\left( t\right) .$ This latter condition, together
with the requirement that the metric at $R=1$ be continuous with the
exterior Schwarzschild metric, led OS to put%
\begin{equation}
t\left( r,R\right) =-\frac{2}{3}y^{3/2}-2\sqrt{y}+\ln \frac{\sqrt{y}+1}{%
\sqrt{y}-1}\quad ,  \label{tyrel}
\end{equation}%
where the \emph{cotime }$y$ is%
\begin{equation}
y\left( r,R\right) =\frac{r}{R}+\frac{R^{2}-1}{2}\quad .  \label{OSy}
\end{equation}%
Note that, while they referred to $t$ as the "external time", this latter
equation is an analytic continuation of $t$ which establishes $\left(
r,t\right) $ as both exterior and interior coordinates. It is now a simple
conclusion, which for some reason OS failed to draw, that the stable end
state, corresponding to $t=+\infty ,$ is obtained by putting $y=1$, that is%
\begin{equation}
r_{\infty }=\frac{R\left( 3-R^{2}\right) }{2}\quad .
\end{equation}

Such a stable end state is in clear contradiction with the conclusion of
Oppenheimer and Volkoff\cite{oppvol} and therefore with the opening sentence
of the OS article itself. In contrast with black hole theory it gives a
material density spread over the entire interior of the "event horizon".
They showed elsewhere in the same article that the initial state ($%
t\rightarrow -\infty $) has a uniform distribution of stellar material; this
means, for example, that half of the total material is contained between $%
R=0.7937$ and $R=1$. Note that equation (\ref{OSy}) informs us that $r\sim R$
as $t$ tends to minus infinity. Thus the OS collapsar has an end state with
half of its material concentrated in a shell between $r_{\infty }=0.9406$
and $r_{\infty }=1;$ we propose for it the title of the \emph{football
collapsar}.

Actually the concentration of matter in the shell of the football is greater
than this calculation indicates. The derivative of $r_{\infty }$ with
respect to $R$ contains a factor $\left( 1-R\right) $, which indicates
infinite density as $R$ tends to 1. However, this is not as extreme as the
singularity thought, by black-hole theorists, to lie at $R=0$; rather, the
physical conditions in the shell resemble those in a white dwarf of solar
mass; a proper treatment of this region will require us to rehabilitate the
notion of a gravitational force, and to incorporate both that force, which
compresses stellar material into the shell, and the degenerate electron
pressure of that material, which resists compression. That is a study which
goes beyond the OS dust model, and which will require a different treatment.

\section{Modifying OS to meet a causality condition}

\bigskip \emph{A small piece of quartz is separated from the diamond.}

We have just argued that the OS article does not support black-hole theory.
However, the OS solution is problematic in that, as $t$ approaches plus
infinity, the collapsar reaches a certain stage when light cannot escape
from it. OS gave minimal details to justify this anomaly, and we pointed out%
\cite{causaps} that this feature of their model conflicts with the
requirement of Hilbert causality enunciated in 1917 by Hilbert\cite{hilbert}%
, and revived more recently by Logunov and Mestvirishvili\cite{log2012}. It
requires only a simple correction to the OS metric\cite{causaps}, which
leads to replacing (\ref{OSy}) by%
\begin{equation}
y=\frac{r}{R}-\frac{R^{2}-6R+5}{4}\quad ,  \label{myy}
\end{equation}%
to ensure that radial light rays leaving $R$ = 0 for all $t$ escape from the
collapsar; this removes the problematic feature of the OS solution, which
itself formed the basis for Penrose's black-hole theorem\cite{penrose}\emph{%
. }The modification \ also increases the shell concentration of the
football, because%
\begin{equation}
r_{\infty }=\frac{R\left( 3-R\right) ^{2}}{4}\quad ,
\end{equation}%
which means that half of the material is now in $0.9659<R<1$.

Setting aside any predilection one might have for the geometrization of
space-time, it is only natural to enquire "What pressure inflates the
football?" In our dust collapsar there are no forces other than gravity, so
once we have asked the forbidden question the only possible answer to it is
that gravity itself, now turned from attraction to repulsion, inflates the
football. A clear way to see repulsive gravity at work is to explore the
trajectory of a test particle which crashes into the surface $R=1$ at
surface radius $r_{0}$, with a speed greater than the speed at which the
surface itself is contracting. We limit attention to radial trajectories for
which the OS metric may be simplified, putting $x=r/R$, to%
\begin{equation}
ds^{2}=xdx^{2}-x^{2}dR^{2}\quad ,
\end{equation}%
for which there is an integral%
\begin{equation}
x^{2}\frac{dR}{ds}=-C\quad \left( C>0\right) \quad .
\end{equation}%
It then follows, putting $x_{0}=r_{0},$ that%
\begin{equation}
R\left( x\right) =1-\int_{x}^{x_{0}}\frac{C}{\sqrt{x^{\prime
3}+C^{2}x^{\prime }}}dx^{\prime }\quad ,
\end{equation}%
and hence, putting $y=1$ in (\ref{myy}), 
\begin{equation}
r_{\infty }=x_{\infty }\left( 3-2\sqrt{x_{\infty }}\right) ,\quad \sqrt{%
x_{\infty }}=1+\frac{1}{2}\int_{x_{\infty }}^{x_{0}}\frac{C}{\sqrt{%
x^{3}+C^{2}x}}dx\quad .
\end{equation}%
We may thus compute the end point of the crash particle as a function of its
time of arrival at the surface (effectively $x_{0}$) and its crash speed
(effectively $C$ ). For the ultrarelativistic case, $C\rightarrow +\infty ,$
the integration comes out in elementary functions and $r_{\infty }$ takes
the value%
\begin{equation}
r_{\infty }^{U}=\frac{1}{4}\left( 1+\sqrt{x_{0}}\right) ^{2}\left( 2-\sqrt{%
x_{0}}\right) \quad ,
\end{equation}%
and, for all finite $C,$ $r_{\infty }>r_{\infty }^{U}$, so that $r_{\infty
}^{U}$ gives the maximum penetration of the football by a crash particle.
Putting $r_{0}=x_{0}=1.45$, which corresponds to the collapsar being only
just inside its own photonsphere at $r=3m,$ gives $r_{\infty }^{U}=0.9666,$
and so, for all radii less than this, the crash particle fails to get beyond
the "half-mass shell" which we obtained at the end of the previous paragraph.

Thus, even ignoring collisions with the material of the football, such a
crash particle encounters a sharp decelerating force on entering the shell,
and that same force is what compresses the stellar material to form the
shell.
\section{A supermassive white dwarf}

 \emph{The facets of the diamond are cut and polished to reveal an
object of wonderful simplicity.}

White dwarfs of around solar mass have been understood from the 1930s in
terms of the Chandrasekhar\cite{chandra} analysis, which (see, for example, 
\cite{weinberg}, section 11.3) balances the Fermi pressure of a degenerate
electron gas against the gravitational force. Although gravity in such a
white dwarf is very strong, and may indeed eclipse the Fermi pressure, it
nevertheless remains Newtonian; the equations which determine the pressure profile are
\begin{eqnarray}
\frac{dp}{dr}&=&-\frac{GM\rho}{r^2} \quad,\nonumber\\
M&=&\int_0^r4\pi \rho(r') r'^2dr\quad,
\end{eqnarray}
together with the equation of state $\rho=\rho(p)$.		

An extension to the strong-gravity regime was made by Oppenheimer and Volkoff (OV)\cite%
{oppvol},  by considering the GR field equations with the
metric%
\begin{equation}
ds^{2}=B(r)dt^{2}-A(r)dr^{2}-r^{2}(d\theta ^{2}+\sin ^{2}\theta d\phi
^{2})\quad ,
\end{equation}%
and the stress tensor
\begin{equation}
T^{\mu\nu}=\left(\rho+\frac{p}{c^2}\right)\frac{dx^\mu}{ds}\frac{dx^\nu}{ds}-\frac{p}{c^2}g^{\mu\nu}\quad.
\end{equation}
These lead to the field equations
\begin{eqnarray}
\frac{du}{dr} &=&4\pi r^{2}\rho \quad ,  \nonumber \\
\frac{dp}{dr} &=&-\frac{G(
\rho +p/c^2) ( u+4\pi r^{3}p/c^2) }{r( r-2uG/c^2) }\quad .
\label{fieldeq}
\end{eqnarray}%
Together with the equation of state, and with suitable initial conditions, these determine the pressure profile $p(r)$ as well as  $u(r)$; the metric coefficients are then given(\cite{weinberg} eqn.(11.1.16)), using units with $G=c=1$, by
\begin{eqnarray}
A &=&\frac{r}{r-2u( r) }\quad
,  \nonumber \\
AB &=&\exp \left[ -\int_{r}^{\infty }\frac{8\pi r'^{2}[ \rho (r') +p(r') ] }{r'-2u( r') }dr'\right] \quad .
\label{AB}
\end{eqnarray}%

The OV article was given little attention until the early 1960s, when it became the basis for the  argument that, above a few solar masses, any collapsar must end up as a black hole\cite{htww}. This was a historic error, occasioned by the failure of Oppenheimer and Snyder (OS)\cite{oppsny} to recognize that their article, written a few months after OV, was in contradiction with the conclusions of OV, and very likely compounded by the twenty year period of obscurity suffered by both articles.

In their 1960s form, the OV field equations (\ref{fieldeq}) were written with $u(r)$ replaced by $M(r)$, and the notation remains unchanged up to the present. Of course, in the Newtonian theory $M(r)$ is the mass inside a sphere of radius $r$, and we arrive at precisely that same understanding if we integrate (\ref{fieldeq}) starting from $u(0)=0$, and then interpret $\rho(r)$ as the mass density.  Weinberg ( \cite{weinberg} section 11.1), while stating explicitly that this latter identification is
incorrect, also states that $u(0)\neq 0$ makes $A(
r)$ singular at $r=0$. In fact $A(0)=0$ in this case, and it is  $B( r)$ which has a pole at $r=0$, a feature which OV had recognized, but which they viewed as no obstacle to a stable solution; in this they were correct, as we shall now demonstrate.

OV considered that $u(0)$ could be zero or negative, but then gave a "disproof" of the latter possibility, by considering, in their footnote 10, a restricted range of possible equations of state. They then showed that, with $u(0)=0$, the pressure increases monotonically from surface to centre, that is gravity remains attractive throughout the interior; then it follows, as for Newtonian white dwarfs, that beyond a limiting mass not much larger than that of our Sun no stable solution is possible. However, as we discussed in Section 2, the OS article shows that, even in the extreme case $p(r)=0$, there is a stable solution for any mass, and such solutions have repulsive gravity in a central region. This means we must reexamine the OV model with $u(0)<0$, for which we find immediately that $p(0)=\rho(0)=0$; so $p(r)$ is initially increasing and this does indeed imply repulsive gravity.

For the equation of state we shall assume that there is a
range of $p$, corresponding roughly to the shell of the OS football, in
which the stellar material is in a state of compression like that in a lower-mass
solar white dwarf, that is a 5/3 isotrope%
\begin{equation}
\rho =k_1p^{3/5}\quad (p\quad \rm{large})\quad,
\end{equation}%
where
\begin{equation}
k_1=\frac{m_D}{3\pi^2}\left(\frac{60\pi^4m_e}{h^2}\right)^{3/5}=3.426.10^{-8}\quad(\rm{in\quad cgs\quad units})\quad,
\end{equation}
$m_e$ being the electron rest mass and $m_D$ the nucleonic mass per electron, which is taken as the neutron mass times\footnote{An atom of iron, the main constituent of a white dwarf, contains 56 nucleons and 26 electrons.} 56/26. 
Near the centre this equation of state is not possible, so we substitute the only one permitted under the OV analysis, namely 
\begin{equation}
\rho =kp\quad (p\quad \rm{small})\quad ,
\end{equation}%
The two ranges of $p$ may
then be combined by putting, for example%
\begin{equation}
\rho =\frac{k_1p^{3/5}\sinh(p/p_1)+kp}{\cosh(p/p_1)}\quad.
\end{equation}%
The parameter $k$ must be greater than 3, this being the extreme value corresponding to a relativistic gas. The integration of (\ref{fieldeq}) may be started analytically from $r=0$, and we may choose units
so that $k_1=1$, in addition to the choices $G=c=1$ which we
made already. Then the initial values of $p$ and $u$ are%
\begin{equation}
p=ar^{(k+1)/2},\quad u=u_0+\frac{8k\pi a}{k+7}r^{(k+7)/2}\quad ,
\label{EoS}
\end{equation}%
with $a$ constant, and from these the integration may be continued numerically,
starting from a suitably small $r$. The  units are\footnote{The OV units were different, because the particles exerting pressure were neutrons, as opposed to the electron gas in a stationary nucleon background which we consider here}
\begin{eqnarray}
\rm{mass\quad unit}&=&6.47.10^{35}\rm{gm}\quad,\\
\rm{length\quad unit}&=&4.79.10^7\rm{cm}\quad,\\
\rm{time\quad unit}&=&1.60.10^{-3}\rm{sec}\quad.
\end{eqnarray}

The parameter $a$ is not arbitrary; it must be adjusted so that the solution satisfies Einstein's requirement \cite{einstpe} that the gravitational and inertial masses be equal. The limit $r\rightarrow\infty$ of $u(r)$ is the gravitational mass of the collapsar, denoted by $M$, and the inertial mass is obtained by integrating the total energy density
 over three-dimensional space. For this quantity Landau and Lifshitz\cite{lanlif} give \footnote{This expression for the inertial mass is not a 4-scalar, and it refers only to a constant field with spherical symmetry. However, it is a 3-scalar, that is an invariant under a transformation $r=f(r_1)$; in particular it retains the same form in either the isotropic or the harmonic frame.}
\begin{equation}
M_{inertial}=\int(2T^0_0-T)\sqrt{-g}drd\theta d\phi=\int_0^\infty 4\pi r^2\sqrt{AB}(\rho+3p)dr\quad,
\label{inertmass}\quad,
\end{equation}
the quantity $AB$ being given by (\ref{AB}). Then, equating these latter two expressions, we obtain a relation between $a$ and $M$, and hence we obtain a collapsar whose density profile, including in particular its overall radius, is a function of just the two equation-of-state parameters $( k,p_1)$.

\section{Conclusion}
We saw in sections 2 and 3 that a collapsar in the strong gravitational field regime can have density decreasing towards its centre; the concentration of matter towards the surface shell led us to call it the "football collapsar". Section 4 checks that same topology is maintained when an electron gas is incorporated in the model, which validates our supermassive white dwarf (SWD) concept. The SWD has relativistic gravity as distinct from the Newtonian gravity of a normal white dwarf (\cite{schutz}, Investigation 12.4) with its density around 1 tonne/cm$^3$ and polytropic index 5/3. The higher density white-dwarf regime (\cite{schutz}, Investigation 12.5) has polytropic index 4/3 and the density profiles of solutions using this are of similar shape. The OS solution of 1939 did have the football profile, but this went unrecognised. Oppenheimer and Volkoff\cite{oppvol} used polytrope equations-of-state, but failed to discover the SWD solution, as they considered only solutions with central maxima in $p$ and $\rho$.

We found the OS analysis had to be modified to satisfy (Hilbert) causality, ensuring connection by light rays is maintained throughout the collapse. Light rays both penetrate the interior and emerge from it.  This modification eliminates the trapped (null) surface of Penrose, which underpins the event horizon concept. In our solution, this is the gravitational radius.  We have shown that incoming particles reaching this radius are strongly decelerated, as they penetrate the increasingly intense repulsive gravitational field and come to rest within the shell (disregarding collisions with dust comprising the shell).

The radio and infrared emissions from SgrA* are thought to emanate from
gas and dust heated to millions of degrees while falling under the central gravity.
In the SWD model such heating would occur just above the shell surface in a
corona. The corona would be created by the inputs of dust, including gas
evaporated from it, and depend on emission-cooling; these complexities could be
modelled, whereas the alternative “interaction with a massive source of gravity”
being source of the observed emissions is vague and speculative\cite{wheeler}.
The X-ray flaring of SgrA* and other distant galaxy centres has been hypothesised
to arise from the shredding and consuming of a passing star\cite{degenaar}.

In the SWD model, this would need a large concentration of matter to impact
the shell and temporarily destabilise it. For example a stellar mass of white-dwarf density as impactor would be less massive by a factor 10$^4$ than the impacted element of the shell. On the other hand, the expected impact in 2013 on SgrA* of the G2 dust/gas cloud of a few Earth masses would be far too small to perturb
the football shell. With a closest approach some 3000 times the gravitational
radius (0.1 AU) and dimension ˜1AU, it could hardly affect even a large corona
of size ˜0.1AU. If SgrA* possesses a much larger accretion disc, this could absorb
material from the G2 cloud. However, the accretion onto the SWD from
such an accretion disc would be over a longer time-scale and would not give the

predicted fireworks display.

Whether the predicted fireworks happen or not, we propose the SWD model as an alternative to the generally assumed ‘black hole’ model for the galactic centre. The Event Horizon Telescope is planned to probe this at the gravitational scale in the coming
years\cite{gilless}, so should be able to distinguish between the SWD and black-hole models.

\end{document}